\documentclass{svproc}

\usepackage[utf8]{inputenc}
\usepackage[toc,page]{appendix}
\usepackage[utf8]{inputenc}
\usepackage{graphicx}
\usepackage[hscale=0.7,vscale=0.8]{geometry}
\usepackage{amsmath}
\usepackage{amssymb}
\usepackage{verbatim}
\usepackage{soul}
\usepackage{url}
\usepackage{xcolor}
\usepackage{textcomp}
\usepackage{multicol}
\usepackage{subcaption}
\usepackage{wrapfig}
\usepackage{enumerate}

\newcommand{\mysaveA}{\vspace*{-6pt}} % after section

\begin{document}

\title{Verifying Provenance of Digital Media:\\
Why the C2PA Specifications Fall Short\\
}

% {\large \textnormal{\textit {(Findings from an Independent Security Review)}}}

%%%%%%%%%%%%%%%%%%%%%%%%%%%%%%%%%%%%%%%%%%%%%%%%%%%%%%%%%%%%%%%%%%%%%%%%%%%%%%%%%%%%%%%%%%%%%%
\author{Enis Golaszewski\inst{1},
Neal Krawetz\inst{2},
Alan T. Sherman\inst{1},
Edward Zieglar\inst{3},\\
Sai K. Matukumalli\inst{1},
Roberto Yus\inst{1},
Carson L. Kegley\inst{1},
Michael Barthel\inst{1},\\
William Bowman\inst{1},
Bharg Barot\inst{1},
Kaur Kullman\inst{1}
}

\institute{
Cyber Defense Lab, University of Maryland, Baltimore County (UMBC), Baltimore, MD 21250\\
\email{\{golaszewski, sherman, smatuku1, 
ryus, ckegley1, mbarthe1, wbowman4, bhargvb1\}@umbc.edu}
\and
Hacker Factor, Fort Collins, Colorado, 
\email{krawetz26@hackerfactor.com}
\and
National Security Agency, Fort George G. Meade, Maryland,
\email{evziegl@uwe.nsa.gov}
}

\begin{comment}
golaszewski,
sherman,
smatuku1,
ryus,
ckegley1,
mbarthe1,
wbowman4,
bhargvb1,
kak@umbc.edu
krawetz26@hackerfactor.com
evziegl@uwe.nsa.gov
\end{comment}

\maketitle

%\titlerunning{Why C2PA falls short}
%\authorrunning{Golaszewski, et al.}
\pagestyle{myheadings}
\markboth{Why C2PA Falls Short}{Golaszewski, et al.}

%%%%%%%%%%%%%%%%%%%%%%%%%%%%%%%%%%%%%%%%%%%%%%%%%%%%%%%%%%%%%%%%%%%%%%%%%%%%%%

\begin{center}
April 23, 2026
\end{center}
\bigskip

\section*{Executive Summary} %%%%%%%%%%%%%%%%%%%%%%%%%%%%%%%%%%%%%%%%%
{\mysaveA}

The rapid rise of generative AI has made it easy to create convincing fake media at scale. 
In response, an industrial coalition has developed the
\textit{Coalition for Content Provenance and Authenticity (C2PA)},
a system intended to provide verifiable provenance for digital content.
Our research team~\cite{C2PAiacr} conducted the first comprehensive, independent security analysis of C2PA.
Our study includes the first formal-methods analysis of C2PA's core protocols.
We find that the current C2PA specifications fail to achieve their claimed security goals.
Furthermore, they also fail to achieve
key additional goals, which all such provenance systems require for trustworthy deployment.
As a result, C2PA may mislead users, platforms, and policymakers if relied upon prematurely.
C2PA is a promising idea, but it should not yet be relied upon for high-stakes uses 
such as financial disclosures, journalism, or legal evidence.

% is not yet ready for use as a trustworthy foundation for public policy, journalism, or legal evidence.
% ATS: Careful not to overreach on policy recommendations.  On the otherhand, we might combine main technical findings and policy implications.  Would be better to frane in terms of policy implications rather than policy recommendations.
% C2PA does not currently achieve its stated security goals and should not yet be relied upon for high-stakes uses such as financial disclosure, journalism, or legal evidence.
% Don't mnention elections -- don't create the impression that a fixed C2PA will help secure elections

\section*{The Problem of Misinformation} %%%%%%%%%%%%%%%%%%%%%%%%%%%%%%%%%%%%%%%%%
{\mysaveA}

The ubiquity of AI has rapidly lowered the cost of producing persuasive, high-quality synthetic media. 
A single actor can fabricate images that appear to depict real events, real people, and real documents, at a scale and fidelity that strains human judgment and existing forensic tools. 
The harms are well understood: fraud enabled by counterfeit evidence, reputation attacks, and political manipulation through fabricated content.  

\section*{What is C2PA?} %%%%%%%%%%%%%%%%%%%%%%%%%%%%%%%%%%%%%%%%%%%%%%
{\mysaveA}

C2PA aims to provide \textit{provenance} signals about how digital media were created and modified.
C2PA attaches machine-verifiable metadata intended to help audiences and platforms determine where, when, and how an asset was created or edited. 
Such provenance signals are an achievable useful partial step toward helping consumers assess 
the \textit{authenticity} of digital objects (e.g., ``Is the image real versus fake?'').
Provenance describes the history of a file,
whereas authenticity concerns, for example, whether the content truthfully represents real-world events.
C2PA provides {provenance signals}, {not proof of authenticity}.

C2PA is not a government or international standard. 
It is an {industry-led effort} consisting of
the C2PA coalition,
a set of technical specifications,
a broader ecosystem with tools (e.g., claim generators) and services (e.g., timestamping), and
a conformance program intended to certify products (e.g., validators).
Major companies including Adobe, Google, Microsoft, Meta, and Amazon participate and have endorsed its adoption.

\section*{What C2PA Claims to Do} %%%%%%%%%%%%%%%%%%%%%%%%%%%%%%%%%%%%%%%%%
{\mysaveA}

C2PA attaches ``\textit{content credentials'' (Crs)} to digital media. These credentials can describe:
who or what created the content (e.g., a camera or AI system),
when and where it was created, and whether it is original or edited.
These credentials are protected using cryptographic digital signatures, 
similar to those used in secure web communications.
However, C2PA was \textit{not} designed to  determine authenticity.
Despite this limitation, misleading promotional materials (e.g.,~\cite{SlickAd2026}) 
and public statements by C2PA advocates have sometimes overstated what the technology delivers, 
creating a risk that policymakers and the public 
will place more trust in C2PA signals than is warranted.

%\noindent % a new paragraph begins here
The C2PA specifications make only two security claims:
\begin{enumerate}[-]
\item[1.] \textit{Claim integrity}: Conforming validators can detect tampering of the Crs.
\item[2.] \textit{Weak file integrity}: Conforming validators can detect if certain bits of the digital asset were modified.
\end{enumerate}
This second claim is very limited in that it applies only to the bits outside of the ``exclusion range''
(a set of bit positions) and does not apply to the bits in the exclusion range.

As we point out in our study, any such provenance system should also include the following three additional
essential security goals:
\begin{enumerate}[-]
\item[3.] \textit{Timestamp agreement}: The claim generator (e.g., camera) and validator should agree on the timestamp.
\item[4.] \textit{Validator consistency}: Conforming validators (sharing the same specification version and trust lists) should agree on the validity of any given claim.
\item[5.] \textit{Strong file integrity}: Conforming validators should detect if any bits of the digital asset were modified (including non-C2PA metadata).
\end{enumerate}

\section*{C2PA Fails to Deliver on Its Claims} %%%%%%%%%%%%%%%%%%%%%%%%%%%%%%%%%%%%%%%%%
{\mysaveA} % What C2PA Actually Does

Due to vulnerabilities in the specifications (Versions~2.2--2.4), 
weaknesses in the cryptographic engineering, and 
limitations of the conformance program, 
the C2PA specifications and implementations do not achieve any
of their claimed security goals or any of the essential security goals:
\begin{itemize}
\item Conforming validators are not required to check for revoked certificates, allowing adversaries to
use compromised keys without detection, defeating all five security goals. 
\item Exploiting inadequate cryptographic bindings, adversaries can replace timestamps without detection.
\item  Missing information (e.g., version number), overly liberal policies, and
certification failing to check source code results in
conforming validators that sometimes give 
contradictory results when assessing the provenance of a digital asset. 
\end{itemize}
Although the underlying \textit{cryptography} in C2PA is battle tested, 
its \textit{integration} in the specifications is not and has failed our tests.

\section*{Key Findings} %%%%%%%%%%%%%%%%%%%%%%%%%%%%%%%%%%%%%%%%%
{\mysaveA}

Our study evaluated the C2PA specifications, real-world implementations, 
the conformance program, and validation tools. 
We identified several systemic weaknesses.
Figures~\ref{fig:forged}--\ref{fig:expired} present four concrete examples that illustrate how adversaries can exploit vulnerabilities in the C2PA specifications to create forgeries and sow confusion.

\subsection*{1. Timestamps Can Be Forged or Altered} %%%%%

\begin{figure}[t] %%%%%%%%%%%%%%%%%%% Fig1
    \centering
    \begin{subfigure}{0.35\textwidth}
    \includegraphics[width=\linewidth]{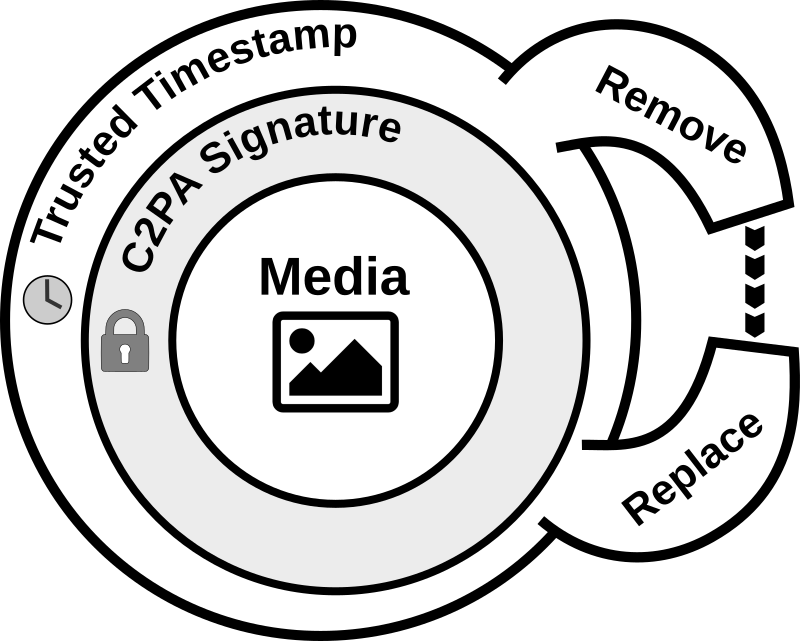}
    \caption{Layers of C2PA protection.}
    \label{fig:C2PA-diagram}
    \end{subfigure}
    \begin{subfigure}{0.6\textwidth}
    \includegraphics[width=\linewidth]{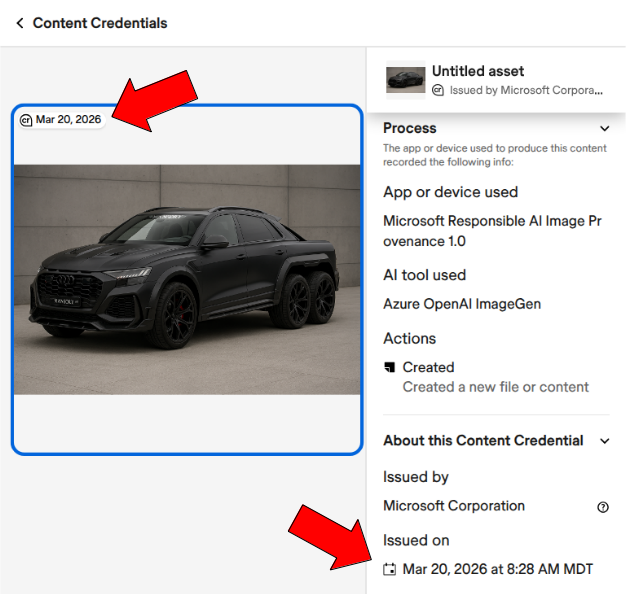}
    \caption{Image with altered timestamp.}
    \label{fig:black-car}
    \end{subfigure}
    \caption{Timestamps can be replaced without detection.
    C2PA uses a layered security approach: the signature protects the media and the optional trusted timestamp protects the signature. Nothing in the signed data references the timestamp, allowing removal and replacement without detection. 
    We altered the timestamp in this image.
    C2PA validators display the date without noting that it may not be original. 
    The arrows point to where the CAI Verify validator displays altered dates.}
    \label{fig:forged}
\end{figure}

%%%% Specify which validator is used here in Fig1

C2PA relies on trusted timestamp authorities to indicate when content was signed. 
However, through our formal-methods analysis, we found that
these timestamps can be replaced or modified without detection. Furthermore, 
conforming validators may accept conflicting timestamps as valid (Figure~\ref{fig:forged}).
This vulnerability undermines trust in determining when content was created---an essential factor 
in journalism and legal contexts.

In our study, we suggest ways using signed timestamped assertions 
to provide evidence when the content was created or signed. 
Our suggestion is not vulnerable to the aforementioned timestamp manipulation.

\subsection*{2. Revoked or Compromised Credentials Are Still Accepted} %%%%%

%\begin{wrapfigure}{l}{0.5\linewidth}
%\end{wrapfigure}

\begin{figure}[t] %%%%%%%%%%%%%%%%%%% Fig2
    \centering
    %\subcaptionbox{SUBCAPTION 1 HERE \label{fig:adobe-horshack}}[0.48\textwidth]
    {\includegraphics[width=0.6\linewidth]{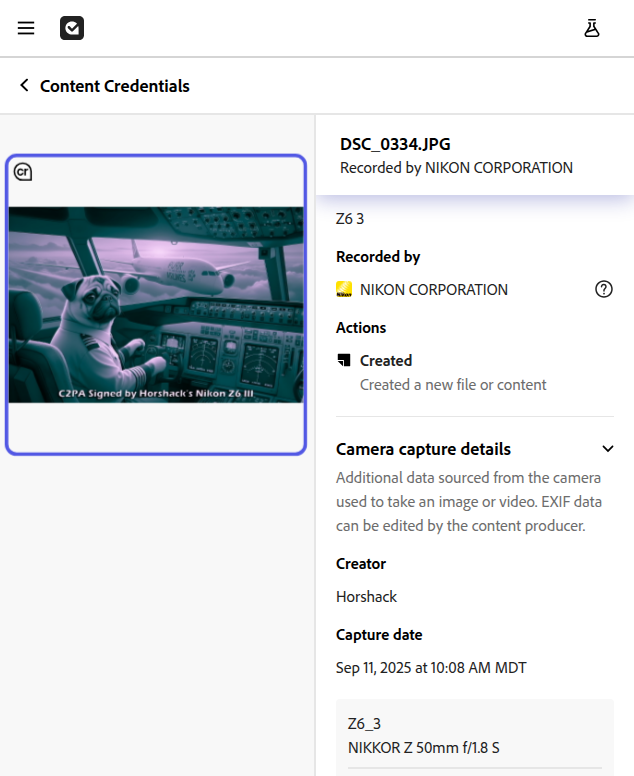}}
    \hfill
    %
    %\subcaptionbox{SUBCAPTION 2 HERE (invalid label enhanced by authors) \label{fig:verifieddit-horshack}}[0.48\textwidth]
    % {\includegraphics[width=\linewidth]{images/verifieddit-horshack.png}}
    \caption{Validators accept revoked certificates and conforming validators present contradictory results, permitting forgeries without detection.
    Using a Nikon Z6 III, Adam Horshack demonstrated how to use the camera to sign an AI image. In November 2025, Nikon revoked the camera's certificate. Over six months later, Adobe Inspect (pictured here) reports the signature as valid, while Verifieddit reports it as invalid. Neither conforming validator reports the revocation.}
    \label{fig:revoked}
\end{figure}

If a signing key is compromised, it should be revoked. However,
(a)~many conforming validators {fail to check revocation properly}, and 
(b)~some conforming validators continue to accept {known compromised credentials}.
This vulnerability allows attackers to create forged content that appears 
legitimate (Figure~\ref{fig:revoked}).

Out of concern for privacy, the C2PA specifications intentionally make checking for revoked certificates optional, and when done, permit such checking only via the \textit{Online Certificate Status Protocol (OCSP)},
expressly forbidding \textit{certificate revocation lists (CRLs)}.
This design choice is flawed in two ways.  First, to achieve the security goals, it is
essential to check for revoked certificates. Second, CRLs offer much better privacy properties
than OCSP provides.

\subsection*{3. Different Tools Produce Contradictory Results} %%%%%

We tested multiple widely used validators and found that
(a)~the same image can be labeled {valid by one tool and invalid by another}, and 
(b)~users may receive {conflicting conclusions} about the same media, including
concerning whether it was AI-generated.
This inconsistency creates confusion and undermines trust (Figure~\ref{fig:revoked}).

\begin{figure}[t] %%%%%%%%%%%%%%%%%%% Fig3
    \centering
    \begin{subfigure}{0.80\textwidth}
        \includegraphics[width=\linewidth]{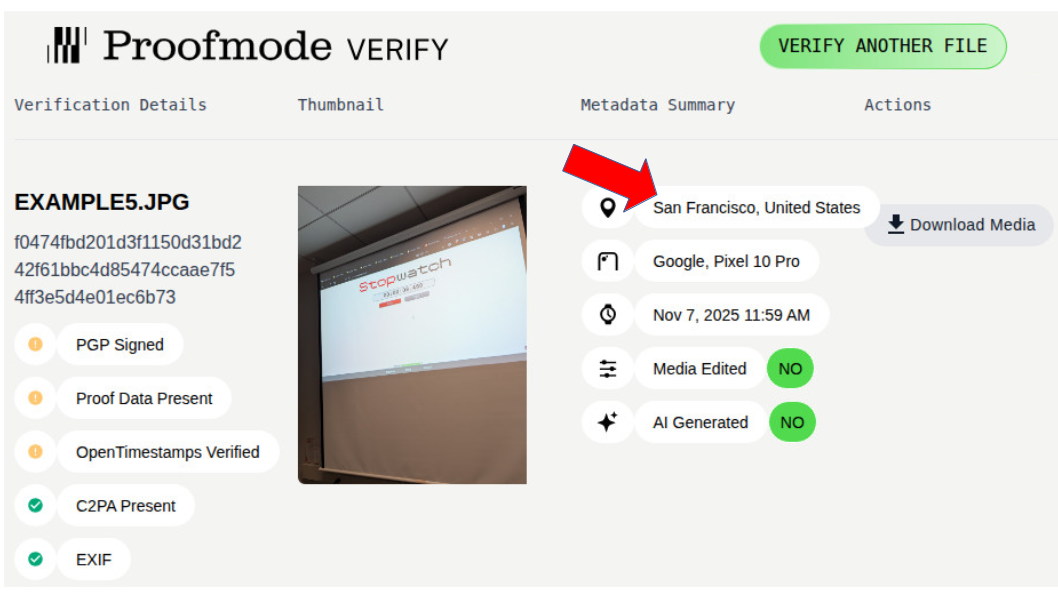}
        \caption{Image with altered GPS location. The arrow points to the modified GPS location.}
        \label{fig:proofmode}
    \end{subfigure}\\[12pt]
    
    \begin{subfigure}{\textwidth}
        \centering
        \includegraphics[width=0.25\linewidth]{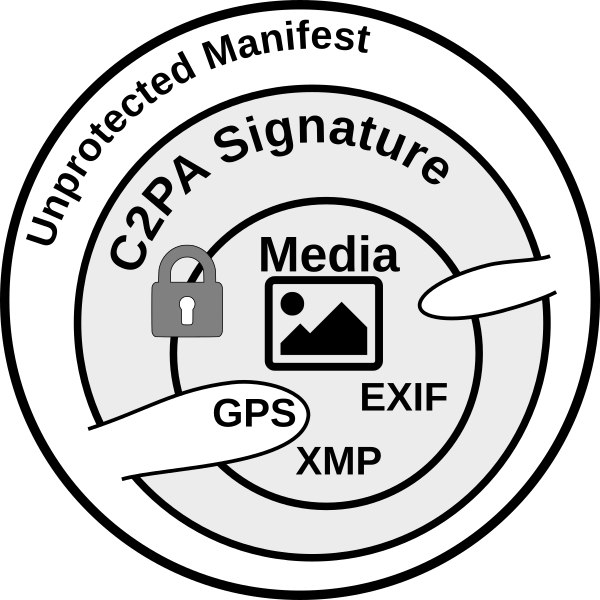}
        \caption{Bytes in the exclusion range can be modified without detection.}
        \label{fig:origin-verify}
    \end{subfigure}
\caption{Google's conforming Pixel~10~Pro camera places GPS information in an exclusion range, enabling an attacker to insert a false GPS location. In this image taken by a Pixel~10~Pro, we modified the GPS location. The conforming Proofmode Verify validator does not detect our alteration and displays the false GPS location.}
        \label{fig:exclusion}
\end{figure} 

\subsection*{4. Parts of Files Can Be Modified Without Detection} %%%%%

The C2PA specifications allow certain parts of a file (such as metadata) to be excluded from 
protection (Figure~\ref{fig:exclusion}).
(a)~These regions can be {altered without invalidating the signature}, and
(b)~important information (e.g., location data) can be changed undetected.
This vulnerability undermines trust.

To address privacy concerns, the C2PA specifications support an exclusion range so that certain
data (e.g., GPS location) may be redacted after the media were created.  In our study, we suggest
a better way (involving additional signatures) to handle such redactions.

\subsection*{5. Credentials Expire and Become Unverifiable}

\begin{figure}[t] %%%%%%%%%%%%%%%%%%% Fig4
    \centering
    \includegraphics[width=0.5\linewidth]{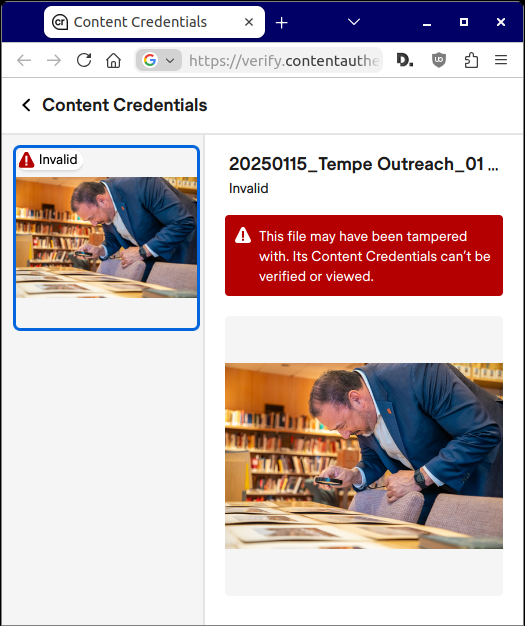}
    \caption{C2PA-signed media can expire and become unverifiable. The image is part of the Arizona Secretary of State's C2PA pilot program. 
    It validated in January 2025, but fails to validate a year later (e.g., by CAI's Verify and Adobe's Content Authenticity Inspect), 
    even though the file has not changed.}
    \label{fig:expired}
\end{figure}

Due in part to expired certificates, some C2PA-signed media have already become unverifiable---sometimes within months (Figure~\ref{fig:expired}). This situation is incompatible with legal
requirements for election record retention (22 months under 52 U.S.C. § 20701), financial records (25 months, 12 C.F.R. § 1002.12), political ads (2 years, 47 C.F.R. § 73.3526), and other archival needs.

\subsection*{6. Certification Does Not Ensure Compliance with the Specifications} %%%%%

C2PA provides a ``conformance program'' to certify products. However,
the conformance program does not define security requirements for conforming validators.
Certification is based largely on self-reported compliance with no examination
of the product's functionality or source code.
As a result, ``conforming'' products do not necessarily comply with the specifications.
Furthermore, because the specifications have vulnerabilities, a product that complies with the
specifications does not necessarily satisfy the claimed security goals.

\section*{Why Our Findings Matter for Policy} %%%%%%%%%%%%%%%%%%%%%%%%%%%%%%%%%%%%%%%%%
{\mysaveA}

C2PA is already being promoted as part of a solution for high-stake applications including:
combating misinformation,
disclosing financial information,
supporting journalism, and
providing legal evidence of digital media.
If the underlying system is unreliable:
false content may be mistakenly trusted,
legitimate content may be wrongly dismissed, and
public confidence may erode further.
Premature adoption of C2PA could {worsen the misinformation problem rather than solve it}.

\section*{Key Recommendations} %%%%%%%%%%%%%%%%%%%%%%%%%%%%%%%%%%%%%%%%%
{\mysaveA}

Before relying on C2PA in policy or regulation, several improvements are essential:
\begin{itemize}
\item {Require strict certificate revocation checking} (including via privacy-preserving methods).
\item {Ensure timestamps are securely bound} to content and cannot be altered without detection.
\item {Mandate consistency across validation tools}.
\item {Protect the entire file} (including non-C2PA metadata), not just selected portions.
\item {Establish independent security audits} for certified products.
\item {Clarify claims and limitations} in public communications about C2PA.
\end{itemize}

Our study offers concrete ways to implement these improvements.
Beyond C2PA, it would be useful to integrate C2PA with services for
reputation, fact-checking, and 
directing users to the original sources of digital assets.

The Pixel~10~Pro and Version~2.3 (January 2026) of the specifications incorporated some of our suggestions.
Version~2.4 (April 2026) does not address any of our concerns.
We hope our remaining recommendations will be implemented by later specification versions and future validators.
C2PA could eventually be positioned to become a widely adopted useful tool 
for helping people to assess the provenance of digital assets.  
We hope that our study and recommendations will accelerate this process and strengthen C2PA.

\section*{Conclusion} %%%%%%%%%%%%%%%%%%%%%%%%%%%%%%%%%%%%%%%%%
{\mysaveA}

C2PA represents an important step toward addressing the growing problem of digital misinformation. 
However, the specifications are currently flawed, incomplete, inconsistent, and vulnerable to misuse.
Policymakers should view C2PA as an {emerging technology}, not a mature solution.
Careful cybersecurity engineering, stronger safeguards, and independent scrutiny are essential before widespread adoption.

%%%%%%%%%%%%%%%%%%%%%%%%%%%%%%%%%%%%%%%%%%%%%%%%%%%%%%%%%%%%%%%%%%%%%%%%%%%%%%

%\newpage
\bibliography{C2PA.bib}
\bibliographystyle{plain}  % ieeetr

\end{document}